\documentstyle[epsfig]{aipproc}

\begin{document}
\title{Frequency Dependent Specific Heat of Amorphous Silica:
A Molecular Dynamics Computer Simulation}

\author{Peter Scheidler, Walter Kob, J\"urgen Horbach, and Kurt Binder}
\address{Institute of Physics, Johannes Gutenberg University,
Staudinger Weg 7, D-55099 Mainz, Germany}

%\lefthead{LEFT head}
%\righthead{RIGHT head}
\maketitle

\begin{abstract}
We use molecular dynamics computer simulations to calculate the
frequency dependence of the specific heat of a SiO$_2$ melt. The ions
interact with the BKS potential and the simulations are done in the
$NVE$ ensemble. We find that the frequency dependence of the specific
heat shows qualitatively the same behavior as the one of structural
quantities, in that at high frequencies a microscopic peak is observed
and at low frequencies an $\alpha$-peak, the location of which quickly
moves to lower frequencies when the temperature is decreased.

\end{abstract}

\section*{Introduction}
The dynamics of supercooled liquids can be studied by many different
techniques, such as light and neutron scattering, dielectric
measurements, NMR, or frequency dependent specific heat measurements,
to name a few~\cite{vigo}. In order to arrive at a better understanding
of these systems also various types of computer simulations have been
used to supplement the experimental data. However, essentially all of
these simulations have focussed on the investigation of {\it static}
properties or have studied the time dependence of {\it structural}
quantities, like the mean squared displacement of a tagged particle or
the decay of the intermediate scattering function.  What these
simulations have not addressed so far, apart from a noticeable study of
Grest and Nagel~\cite{grest87}, is the time dependence of thermodynamic
quantities, like the specific heat. The reason for the lack of
simulations in this direction is that the accurate determination of
this quantity in a simulation is very demanding in computer resources
because of its collective nature. This fact is of course very
regrettable since one of the simplest ways to determine the glass
transition temperature in a real experiment is to measure the (static)
specific heat. Using ac techniques it is today also possible to measure
the {\it frequency dependent} specific heat, $c(\nu)$, and thus to gain
more insight into this observable~\cite{birge85_jeong95}.  What is so
far not possible in real experiments is to measure $c(\nu)$ at
frequencies higher than 1MHz, and thus the influence of the microscopic
dynamics, which is in the THz range, cannot be investigated. For
computer simulations it is, however, no problem to study $c(\nu)$ also
at these high frequencies and in this paper we report the outcome of
such an investigation for the strong glass former silica.

\section*{Model and Details of the Simulation}

The silica model we use is the one proposed by van Beest {\it et
al.}~\cite{beest90}. In this model the interaction $\phi(r_{ij})$
between two particles $i$ and $j$ a distance $r_{ij}$ apart is given 
by a two body potential of the form
\begin{equation}
\phi(r_{ij})=\frac{q_i q_j e^2}{r_{ij}}+A_{ij}\exp(-B_{ij}r_{ij})-
\frac{C_{ij}}{r_{ij}^6}\quad .
\label{eq1}
\end{equation}
The values of the partial charges $q_i$ and the constants $A_{ij}$,
$B_{ij}$ can be found in Ref.~\cite{beest90}. Since the quantity we
want to investigate, $c(\nu)$, is a collective one, it is necessary to
average it over many independent realizations. Thus the system sizes we
used are rather small, 336 ions, despite the fact that the dynamics of
such a small system will show appreciable finite size
effects~\cite{horbach96}. However, exploratory runs with larger systems
showed that these effects do not change the results substantially. The
simulations were done at constant volume using a box size of 16.8~\AA,
thus at a density of 2.36g/cm$^3$, close to the experimental value of
the density, which is at 2.2g/cm$^3$. The equations of motion have been
integrated with the velocity form of the Verlet algorithm with a time
step of 1.6~fs. The temperatures investigated were 6100~K, 4700~K,
4000~K, 3580~K, 3250~K, and 3000~K. At all temperatures the
system was first equilibrated for a time which is significantly longer
than the typical $\alpha$-relaxation time of the system at this
temperature.

\vspace*{-2mm}
\section*{Results}
In real experiments the frequency dependent specific heat is usually
measured in the $NPT$ ensemble. Although algorithms exist with which the
{\it static equilibrium} properties of a system can be measured in a
simulation in this ensemble, these algorithms introduce an artificial
dynamics of the particles and are therefore not suited to investigate
the dynamical properties of the system in this ensemble. Hence we
calculated the frequency dependent specific heat in the microcanonical
ensemble. Whereas in the $NPT$ ensemble the specific heat is related to
the fluctuations of the enthalpy, in the $NEV$ ensemble it is related to
the fluctuations of the kinetic energy~\cite{grest87,scheidler99}. It
can be shown that in this ensemble the specific heat at frequency $\nu$ is
given by
\begin{equation}
c(\nu)=\frac{k_B}{2/3-K(t=0)-i2\pi\nu\int_0^{\infty}dt\exp(i2\pi\nu t)
K(t)} \quad,
\label{eq2}
\end{equation}
where $K(t)$ is the autocorrelation function of the kinetic energy
$E_{\rm kin}$ and is defined as
\begin{equation}
K(t)=\frac{N}{\bar{E}_{\rm kin}^2}
\left[(E_{\rm kin}(t)-\bar{E}_{\rm kin})
(E_{\rm kin}(0)-\bar{E}_{\rm kin})\right] \quad .
\label{eq3}
\end{equation}
Here $\bar{E}_{\rm kin}$ is the mean kinetic energy and $N$ is the
total number of ions. The derivation of Eq.~(\ref{eq2}) can be found
in Ref.~\cite{scheidler99}.

\begin{figure}[t!] 
\centerline{\epsfig{file=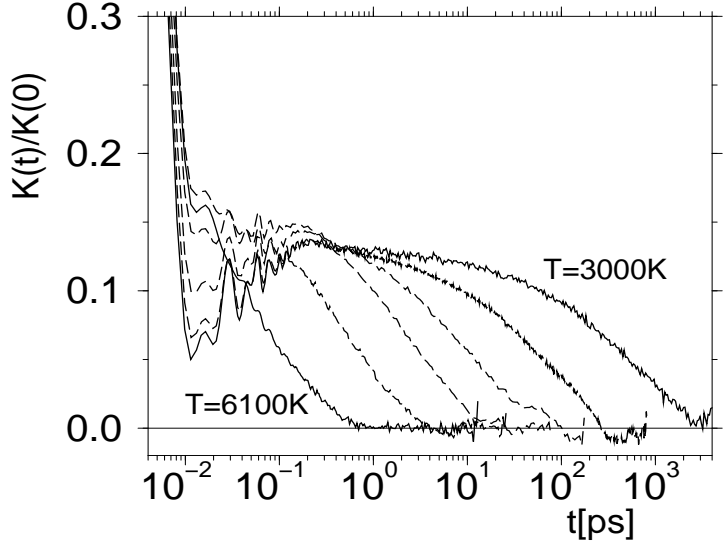,height=2.8in,width=3.8in}}
\vspace{5pt}
\caption{Time dependence of the autocorrelation function of the
kinetic energy for all temperatures investigated.}
\label{fig1}
\end{figure}

From Eq.~(\ref{eq2}) we see that the relevant quantity is the
autocorrelation function $K(t)$. The time dependence of this quantity,
normalized by its value at time $t=0$, is shown in Fig.~\ref{fig1} for
all the temperatures we investigated.

From this figure we recognize that for high temperatures $K(t)/K(0)$
decays very quickly to a value around 0.13 and then goes to zero
like a stretched exponential. With decreasing temperature the
function shows a plateau at intermediate times, the length of which
increases quickly with decreasing temperature. Such a time and
temperature dependence is very similar to the one found for the
relaxation behavior of structural quantities, such as the intermediate
scattering function~\cite{horbach98}. Apart from these features the
curves for the lowest temperatures show also a local minimum at around
0.02~ps, the depth of which increases with decreasing temperatures.
The existence of this dip, as well as the observed high frequency
oscillations, can be understood by realizing that within the harmonic
approximation, which will be valid at even lower temperatures, the
correlator is closely related to the autocorrelation function of the
velocity, which is well known to show such a dip.

\begin{figure}[t!]
\centerline{\epsfig{file=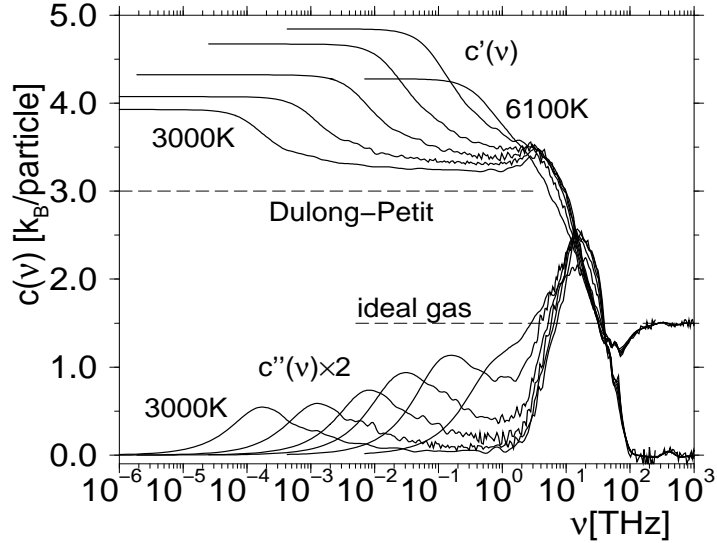,height=2.8in,width=3.8in}}
\vspace{5pt}
\caption{Frequency dependence of the real and imaginary part (times
2.0) of the specific heat for all temperatures investigated. The upper
and lower dashed lines are the values of $c'(\nu)$ for the harmonic
solid and the ideal gas, respectively.}
\label{fig2}
\end{figure}

Using Eq.~(\ref{eq2}) we calculated from $K(t)$ the frequency dependent
specific heat $c(\nu)$. The real and imaginary part of this quantity
are shown in Fig.~\ref{fig2} for all temperatures investigated. Let us
first discuss $c'(\nu)$: For very high frequencies we expect this
function to go to the ideal gas value of 1.5, since the configurational
degrees of freedom are not able to take up energy at such high
frequencies. With decreasing $\nu$ the function shows a fast increase
in the frequency range which corresponds to the microscopic vibrations.
For low temperatures this regime is followed by a plateau the height of
which corresponds to the static specific heat of the system {\it if no
relaxation would take place}, i.e. to the specific heat of the
vibrational degrees of freedom. Since, however, on the time scale of
the $\alpha$-relaxation time the system relaxes, $c'(\nu)$ shows at the
corresponding frequencies a further upward step. This feature is
related to the fact that for very long times, or small frequencies,
those configurational degrees of freedom which are not of vibrational
type are relaxing and thus can take up energy. At even smaller
frequencies the curves then show a plateau, the height of which is the
static specific heat of the system. We see that with decreasing
temperature this value is decreasing but is always significantly above
the harmonic value given by the Dulong-Petit value of 3.0, since the
relaxing configurational degrees of freedom give rise to an enhancement
of the static specific heat.

We also note that the height of the first step in $c'(\nu)$ [coming
from the low frequency side] is the configurational part of the
specific heat. We see that at the lowest temperature this height is
rather small, 0.7$k_B$ per particle, in agreement with the experimental
observation for strong glass-formers.

All these features can also be seen well in the imaginary part of
$c(\nu)$. At high frequencies we have a microscopic peak which
corresponds to the vibrational degrees of freedom. The location and the
height of this peak is essentially independent of temperature. At
intermediate and low temperatures a second peak is seen at low
frequencies, the so-called $\alpha$-peak. Its position depends strongly
on temperature in agreement with the observation that the
$\alpha$-relaxation time of structural quantities increases quickly
with decreasing temperatures~\cite{horbach98}. From the location of
this peak we can read off $\nu_{\rm max}$, the frequency scale of the
relaxation of the specific heat. As it will be shown
elsewhere~\cite{scheidler99}, the product of $\nu_{\rm max}$ with
$\tau(T)$, the $\alpha$-relaxation time of the intermediate scattering
function, is essentially independent of the temperature, thus showing
the intimate connection between the frequency dependent specific heat
and the structural relaxation, in agreement with the prediction of
G\"otze and Latz~\cite{gotze89}.

To summarize we can say that we have presented the results of a large
scale molecular dynamics computer simulation of a realistic model of
viscous silica to investigate the frequency dependence of the specific
heat. In the frequency regime which is accessible also to experiments
our results are in qualitative agreement with the experimental
data~\cite{birge85_jeong95}. At higher frequencies we see the influence
of the vibrational degrees of freedom on $c(\nu)$.  Since no
experimental data for $c(\nu)$ is available for silica we are not able
to compare the results of the present simulations with reality.
However, in a previous investigation we have shown that the present
model gives very good quantitative agreement of the {\it static}
specific heat with the one of real silica~\cite{horbach99} and thus it
is not unreasonable to assume that the results of the simulation on the
dynamic quantity is reliable also.

\vspace*{-2mm}
\section*{ACKNOWLEDGMENTS}
We thank U. Fotheringham for suggesting this work and A. Latz for
many helpful discussions. Part of this work was supported by Schott
Glaswerke, by SFB 262/D1 of the Deutsche Forschungsgemeinschaft, and
BMBF Project 03~N~8008~C.

\end{document}